# HIGH FREQUENCY EPR SPECTROSCOPY OF GERMANIUM CUPRATE.
# I. THE CASE OF DOPING WITH COBALT.

S.V. Demishev[1,4,#], A.V. Semeno[1], N.E. Sluchanko[1], N.A. Samarin[1], A.A. Pronin[1], Y. Inagaki[2], S Okubo[2], H. Ohta[2], Y. Oshima[2], L.I. Leonyuk[3,♣].

[1] *A.M.Prokhorov General Physics Institute of Russian Academy of Sciences, Vavilov street, 38, 119991 Moscow, Russia*
[2] *Molecular Photoscience Research Center, Kobe University, 1-1 Rokkodai, Nada, Kobe 657-8501, Japan*
[3] *M.V.Lomonosov Moscow State University, Vorob'evy Gory, 119899 Moscow, Russia*
[4] *Moscow Institute of Physics and Technology, Dolgoprudny, 141700 Moscow region, Russia*

**Abstract.**

Resonant magnetoabsorption spectra for $CuGeO_3$ single crystals containing 2% of Co impurity are studied in the frequency range 60-360 GHz in magnetic field up to 16 T at temperatures 2-60 K for the case when magnetic field ***B*** is aligned along ***a*** crystallographic axis. It is found that for ***B***||***a*** together with the resonance on $Cu^{2+}$ chains a new, unknown before for doped $CuGeO_3$, absorption line related with $Co^{2+}$ ions appears in EPR spectrum. The *g*-factor magnitudes are $g \approx 2.15$ and $g \approx 4.7$ for the $Cu^{2+}$ chains and $Co^{2+}$ ions respectively. The observed resonances demonstrate difference in the temperature dependence of the line width: while line width for the resonance on $Co^{2+}$ ions decreases with lowering temperature in agreement with the standard theories of spin relaxation, the EPR line width for $Cu^{2+}$ chains increases three times when temperature is decreased from 60 K to 4.2 K. The temperature dependence of magnetisation for $Cu^{2+}$ chains calculated from the integrated intensity data shows two characteristic features, namely the spin-Peierls transition at $T_{SP}=12$ K and a kink at $T=35$ K. Analysis indicates that for the explanation of the obtained data it is necessary to assume that about 10% of $Cu^{2+}$ chains undergo spin-Peierls transition, whereas the rest 90% of the chains are characterized by complete damping of the spin-Peierls state caused by doping with cobalt impurity. At the same time no sign of the antiferromagnetic transition in the range $T>2$ K have been observed for either subsystem of chains $Cu^{2+}$, or subsystem of $Co^{2+}$ ions. In conditions of damping of the spin-Peierls state for the majority of the $Cu^{2+}$ chains the temperature dependence of magnetisation is close to Curie law, whereas for $Co^{2+}$ ions the magnetisation is described by Curie-Weiss law with negative characteristic temperature $\Theta = -(1.8 \pm 0.5)$ K. The obtained results exhibit considerable deviations from the universal scenario of doping for $CuGeO_3$ and discussed in the frames of an alternative theoretical approaches: quantum critical behavior accounting the EPR theory for one-dimensional systems and the three-dimensional antiferromagnet with the reduced by disorder Neel temperature.

---

[#] E-mail: demis@lt.gpi.ru
[♣] Diseased.

# 1. Introduction.

Since the discovery of the inorganic spin-Peierls compound [1] a variety of works focused on effects of doping and its influence on the ground state of this low dimensional magnet have been carried out. It was found that changes of the physical properties of $CuGeO_3$ caused by such impurities like silicon [2,3], zinc [4-7], magnesium [8,9] and nickel [10-12] could be understood in the frame of the universal temperature-concentration, $T$-$x$, phase diagram, which, in its turn, was in qualitative agreement with some theoretical calculations [13]. The characteristic features of the universal scenario of doping are (i) coexistence of quasi one-dimensional (1D) spin-Peierls state with three-dimensional (3D) Neel state in the range of impurity concentration $x<x_c\sim$2-4% and (ii) rapid decrease of the transition temperature into spin-Peierls state $T_{SP}(x)$ when concentration of impurity $x$ is increased. In the region $x>x_c$ the complete damping of the spin-Peierls state occur and only transition into antiferromagnetic phase could be observed.

From the experimental point of view the aforementioned result have been obtained from the study of the both static and dynamic magnetic properties of the doped $CuGeO_3$. In the coexistence region $x<x_c$ the electron paramagnetic resonance (EPR) mode in the $S$=½ antiferromagnetic spin chains appears together with the antiferromagnetic resonance (AFMR) modes corresponding to 3D Neel state [2,7,9,11]. Interesting, that in all known cases no EPR lines belonging to doping impurities in $CuGeO_3$ matrix have been observed. Therefore an important characteristic of the universal doping mechanism is the loose of the "individuality" of impurity, and doping manifests itself only via modification of the properties of $Cu^{2+}$ chains. It is worth to note that this "rule" have worked for impurities which substitute germanium (for instance, silicon), as well as for impurities Zn, Mg, Ni which substitute cooper. In addition the universal doping mechanism is believed to be valid for both magnetic (Ni) [10-12] and non-magnetic (Si, Zn, Mg) [2-9] impurities.

At the same time even first publications on the doping problem have mentioned specific behavior of various impurities in doped $CuGeO_3$. In particular, a considerable deviation from the standard universal scenario of doping has been reported for manganese impurity [14]. More recent studies have revealed another cases of anomalous performance. For example the $CuGeO_3$ samples doped with Mg have shown the re-entrance to spin-Peierls state in vicinity of the critical concentration $x_c$ [15], which probably could not be explained in the theoretical approach [13]. It has been discovered in Ref. 16,17 that doping of $CuGeO_3$ with magnetic impurity of iron on the level $x$=1% leads to damping of both spin-Peierls and antiferromagnetic transitions and onset of the quantum critical behavior. In this case the ground state for $T\leq30K$ is represented by Griffiths phase, which is characterized by divergent magnetic susceptibility described by power law $\chi(T) \propto 1/T^{0.35}$ [18,19]. The above result probably indicates that depending on the chemical nature of impurity the same doping level may correspond to a weak disorder limit, for which the universal scenario is valid [13], or to a strong disorder limit, which is characterized by complete destruction of the long range magnetic order [16,17,20].

As long as the most pronounced deviations from the universal behavior have been found for Mn [14] and Fe [16,17,20] impurities, it is worth to investigate effect of doping by different magnetic impurities on the physical properties and ground state of $CuGeO_3$. The aim of the present work consists in the studying of $CuGeO_3$ doped by cobalt by means of high-frequency ($\omega/2\pi > 60$ GHz) electron paramagnetic resonance (EPR). The choice of the samples with Co impurity has been caused by several reasons. First of all, according to Ref. [21] this magnetic impurity induces a strong Curie-like contribution in magnetic susceptibility and possible spin-Peierls features are observed on the dominating paramagnetic background. Consequently in the EPR spectra additional lines caused by absorption on Co ions in $CuGeO_3$ matrix may be expected. Note that separate signal from the doping impurity in $CuGeO_3$ have not been observed so far. Secondly, it is interesting to compare the influence of the different impurities from the iron group substituting cooper in $Cu^{2+}$ chains and having different spin values: $S$=1 ($Ni^{2+}$), $S$=3/2 ($Co^{2+}$) and $S=2$ ($Fe^{2+}$). Thirdly, as far as we know, the information about EPR spectra of $CuGeO_3$ doped with Co is missing.

**2. Experimental technique.**

Synthesis of the $CuGeO_3$:Co single crystals with impurity concentration $x_{Co}=2\%$ have been carried out using self-flux technique [21,22]. The quality of samples was controlled by X-ray analysis and Raman scattering data. The actual content of impurity in the sample was determined by chemical analysis. We find that within experimental error the actual concentration of impurity have coincided with the nominal (embedded in the charge) one. This result agrees with the data reported in [21], according to which the considered concentration lies below the solubility limit of the Co impurity in $CuGeO_3$. The choice of the material containing 2% of Co for detail investigation have been based on the fact [21] that for $x_{Co}=2\%$ a practically complete damping of the spin-Peierls transition and rise of the strong paramagnetic background occur.

Studying of the resonant magnetoabsorption spectra of microwave radiation for $CuGeO_3$:Co was performed by $\omega/2\pi=60$ GHz cavity EPR spectrometer. For higher frequencies $\omega/2\pi=100$-360 GHz a quasi-optical technique have been used. In the latter case the transmission through the sample have been recorded as a function of magnetic field at fixed microwave radiation frequency. As well as in cavity experiments in quasi-optical measurements together with the spectrum from the sample a reference spectrum from DPPH have been registered. All measurements of the EPR spectra for $CuGeO_3$:Co corresponded to magnetic field ***B*** located along ***a*** crystallographic axis.

**3. Experimental results.**

*3.1 EPR spectra, g-factors and line widths.*

Experimental spectrum of the resonant magnetoabsorption in $CuGeO_3$:Co is formed by two broad lines, which can be completely resolved for frequencies $\omega/2\pi>100$ GHz (an example of the recorded spectrum at $T=4.2$K is shown in fig. 1,a). It is found that frequencies $\omega_{1,2}$ of modes 1 and 2 depends linearly on the resonant magnetic field $B_{res}$: $\omega_{1,2} \sim B_{res}$ (fig 1,b). This behavior suggests that neither resonance 1 nor resonance 2 may be connected with antiferromagnetic resonance expected in the standard co-existence mechanism [13] as long as for AFMR modes in $CuGeO_3$ the dispersion curves $\omega(B_{res})$ are essentially non-linear and $\omega(B_{res}=0)\neq 0$ [2,7,11].

In addition to studying of the spectra for various frequencies at liquid helium temperatures we have investigated influence of temperature on parameters of the lines 1 and 2 taken at 315 GHz. It is visible from figure 2 that in case of resonance 1 the lowering temperature leads to decrease of the resonant field accompanied with the strong broadening of the absorption line. For all temperature studied the resonance 1 has lorentzian shape and fitting of experimental data by lorentzian function in the interval $2 \leq T \leq 60$ K (see solid lines in fig. 2) have allowed to calculate temperature dependences of the line width $w(T)$ resonant field $B_{res}(T)$ and integrated intensity $I(T)$.

The resonance 2 at frequency 315 GHz is characterized by smaller amplitude and consequently by more poor signal to noise ratio (see inset in fig. 2). As a result the accuracy of fitting of line 2 by model Lorentzian becomes insufficient and $I(T)$ data in this case have been obtained by direct integration of the experimental curve. Than we have estimated $w(T)$ assuming that this line have Lorentzian shape. It is worth to note that due to considerable line width for the resonance 2 the accuracy of determination of the resonant field this case have not allowed to discuss temperature dependence of this parameter (see inset in fig. 2). Additionally the temperature range where line 2 was detected has been much smaller than for the resonance 1 and was limited by domain $2 \leq T \leq 20$ K.

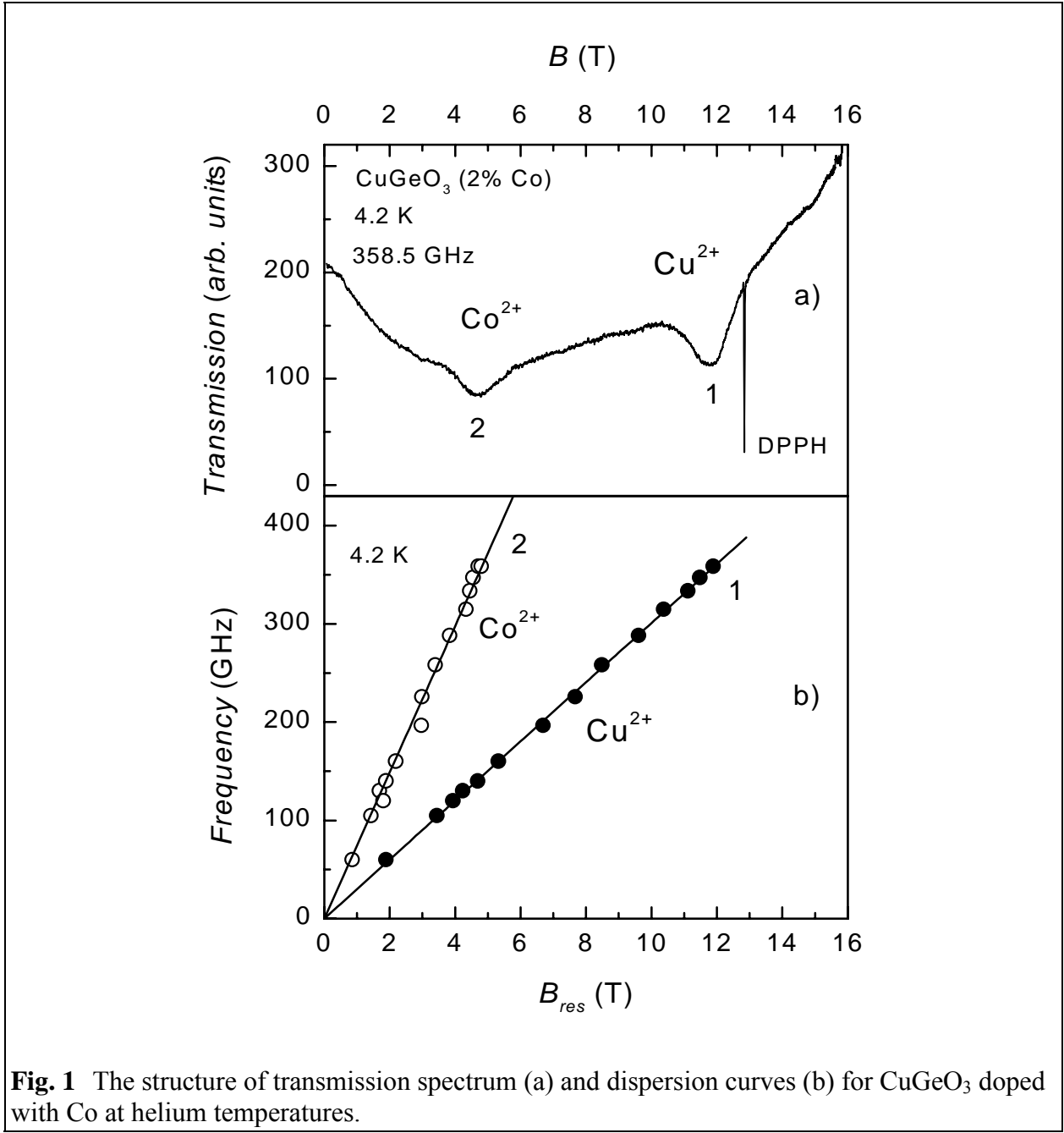

**Fig. 1** The structure of transmission spectrum (a) and dispersion curves (b) for CuGeO$_3$ doped with Co at helium temperatures.

Following the fact that dispersion curves demonstrate an EPR-like behavior $\omega_{1,2} \sim B_{res}$ for the identification of the resonances it is possible to consider corresponding to them $g$-factors (fig. 3). It is visible that for the resonance 1 $g$-factor is close to the value $g \approx 2.15$ characteristic for the EPR on the Cu$^{2+}$ chains measured in the geometry $\boldsymbol{B}\|\mathbf{a}$ [2]. A Lorentzian line shape, which is known to be a fingerprint of the EPR on Cu$^{2+}$ chains in CuGeO$_3$ [23,24,16,17], may serve as an additional argument favoring proposed interpretation.

For the resonance 2 the observed $g$-factor is about two times bigger than in case of resonance 1 and reaches value $g \approx 4.7 \pm 0.2$ (fig. 3). It is known that cobalt impurity substitute cooper in CuGeO$_3$ [21] and the ion Co$^{2+}$ for the certain types of crystal field symmetry may have big $g$-factor magnitudes up to $g \approx 4.3$ [25]. Therefore we suppose that this feature in the resonant magnetoabsorption spectrum may be caused by EPR of impurity ions Co$^{2+}$ in CuGeO$_3$ matrix. An enhancement of the $g$-factor from $g \approx 4.3$ to the value $g \approx 4.7$ may probably reflect either effects of splitting in the crystal field or presence of the strong interactions in the spin clusters suggested in [11] for the doped germanium cuprate. Alternative explanations of the resonance 2 will be considered below in section 4.

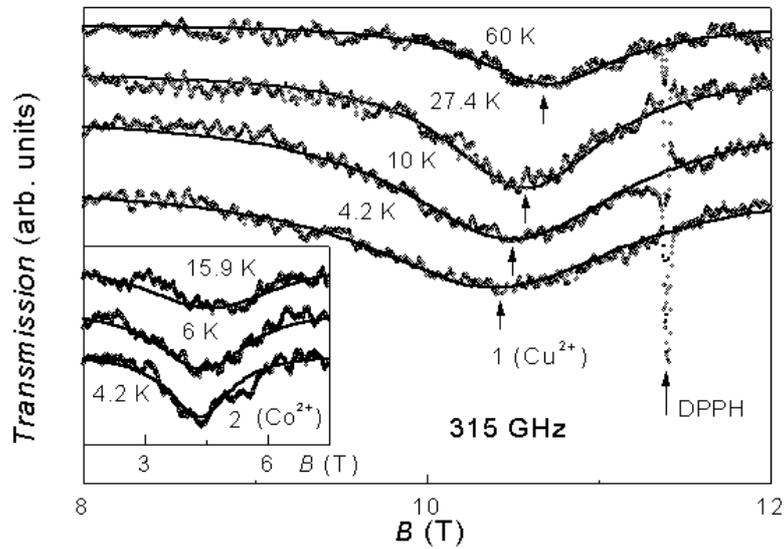

**Fig. 2** The influence of temperature on the resonances 1 ($Cu^{2+}$) and 2 ($Co^{2+}$). Arrows mark the positions of the resonance 1 at various temperatures and DPPH line. Solid lines denote approximations by Lorentzians (see text).

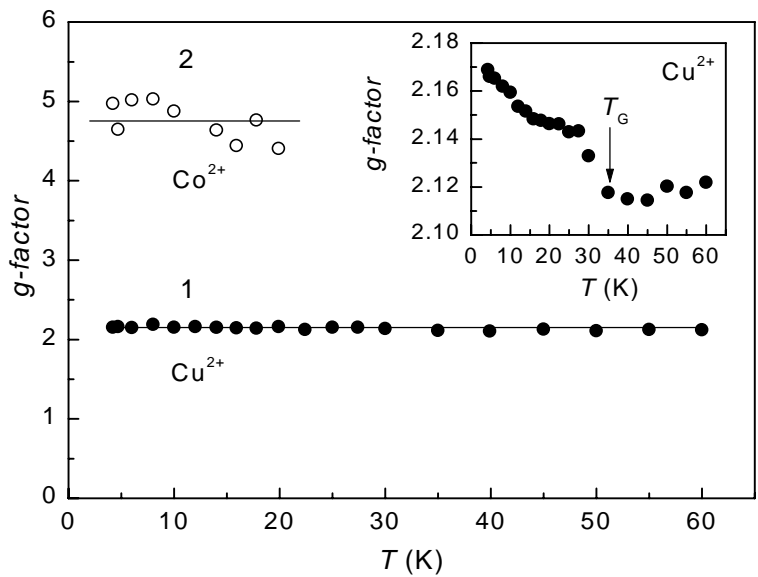

**Fig. 3** The $g$-factors for $CuGeO_3$:Co. The inset shows temperature dependence of the $g$-factor for the resonance on $Cu^{2+}$ chains on enhanced scale.

Interesting, that the temperature dependence of the $g$-factor for $Cu^{2+}$ chains indicates the presence of the temperature region $T \sim T_G$, where their magnetic properties undergo considerable changes (see inset in fig. 3). For $T \geq T_G = 35K$ the $g$-factor weakly depends on temperature:

$g(T)\approx const$, whereas in the interval 30-35 K $g$-factor magnitude increases abruptly. Further decrease of temperature in the range $T<T_G$ causes another section of the $g(T)$ growth (inset in fig. 3).

Different physical nature of the resonances 1 and 2 is appear clearly in the temperature dependences of the line widths $w(T)$ (fig. 4). For the resonance 2 ($Co^{2+}$) the line width increases with temperature in full accordance with the classic theory of spin relaxation [26]. At the same time in the temperature range $T\leq 60K$ the resonance 1 ($Cu^{2+}$) demonstrate an anomalous growth of the $w(T)$ when temperature is lowered. In the case of quasi one-dimensional antiferromagnetic $S=\frac{1}{2}$ spin chains consisting of $Cu^{2+}$ ions the low temperature broadening of the EPR line width may reflect influence of the staggered field [24] or approaching to the antiferromagnetic transition point [27] (both possibilities will be considered in detail in section 4).

Therefore experimental data presented in fig. 1-4 suggest that doping of $CuGeO_3$ with cobalt impurity on the level $x=2\%$ leads to appearance of the new line of absorption probably caused be EPR on $Co^{2+}$ ions embedded in $CuGeO_3$ matrix. Simultaneously the cobalt impurity modifies the properties of the resonance on the $Cu^{2+}$ chains inducing an anomalous temperature dependence of the line width (note that in pure $CuGeO_3$ the EPR line width decreases with lowering temperature [2,7,23]).

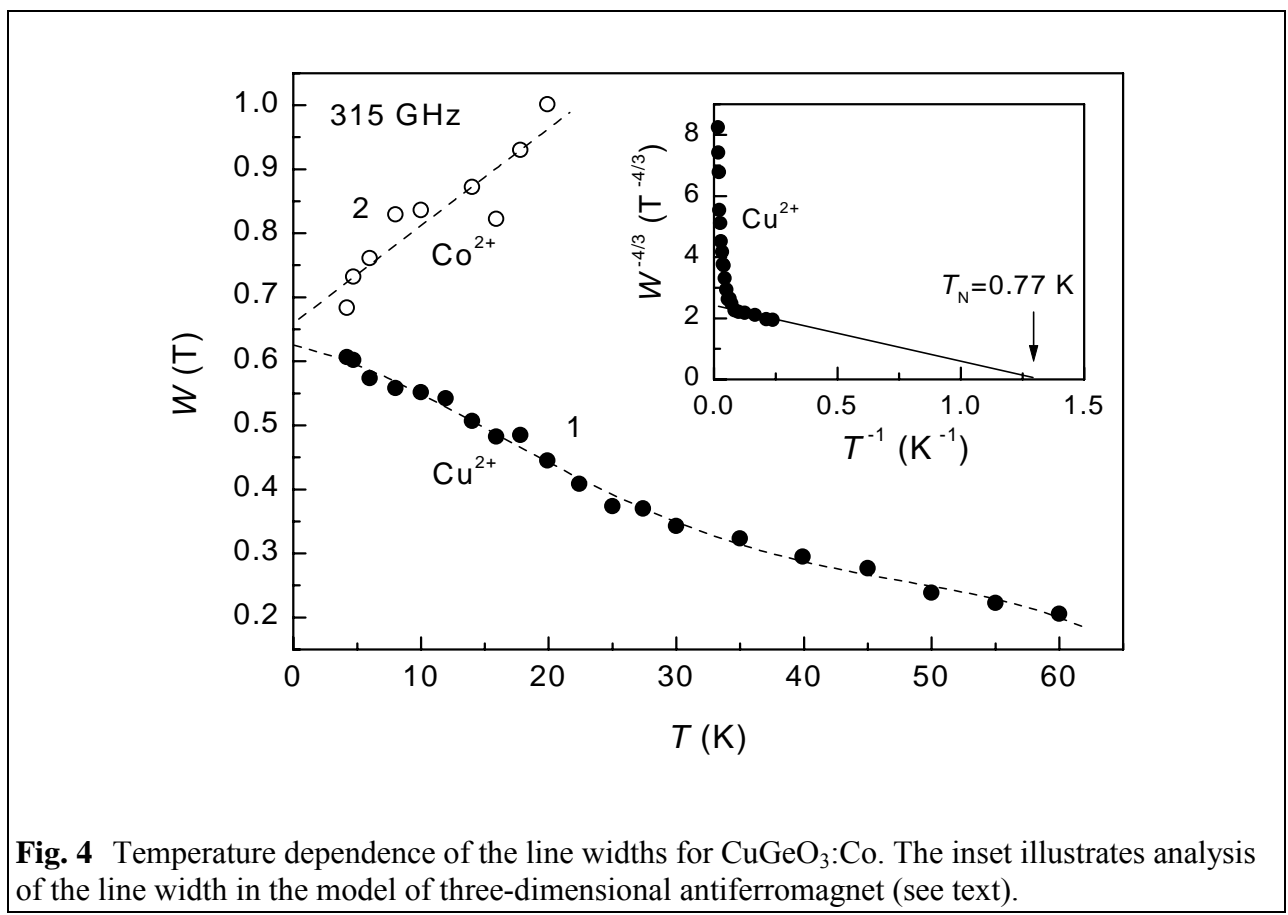

**Fig. 4** Temperature dependence of the line widths for $CuGeO_3$:Co. The inset illustrates analysis of the line width in the model of three-dimensional antiferromagnet (see text).

3.2 *Temperature and field dependencies of the integrated intensity and magnetization.*

The possible EPR nature of the two modes forming resonant magnetoabsorption spectrum in $CuGeO_3$:Co suggests to analyze temperature and frequency (field) dependencies of the integrated intensity. However our experiments have been carried out at high frequencies required for the separation of various spectral components. Therefore it is necessary to take into account possible violation of the linear relation between integrated intensity and magnetic susceptibility $I(T)\sim\chi(T)$

[25,26]. For the arbitrary values of frequency and resonant magnetic field $B_{res}$ the integrated intensity is given by [17]

$$I(T) \sim \omega \frac{M(T, B_{res})}{B_{res}}, \qquad (1)$$

where $M(T, B_{res})$ denotes magnetisation related to EPR mode $\omega(B_{res})$. In quasi-optical experiments in millimeter or submillimeter wavelength range it is convenient to use parameters of the DPPH EPR line as a reference:

$$\frac{I_{1,2}}{I_0} = \frac{M_{1,2}(T, B_{res}^{1,2})}{M_0(T, B_{res}^0)} \cdot \frac{B_{res}^0(\omega)}{B_{res}^{1,2}(\omega)} . \qquad (2)$$

Here indexes 1, 2 and 0 denote integrated intensities of EPR lines for $Cu^{2+}$ chains, $Co^{2+}$ ions and DPPH respectively. As long as DPPH magnetisation can be known from independent measurements, the equation (2) allows to calculate temperature and field dependences $M_1(T,B_{res})$ and $M_2(T,B_{res})$ for $Cu^{2+}$ and $Co^{2+}$ from the ratios $I_{1,2}/I_0$, which can be found directly from the measured resonant magnetoabsorption spectra. Note that the ratio $I_{1,2}/I_0$ weakly depends on the complicated frequency characteristic of the microwave line, which connects oscillator, sample and receiver. Therefore the data processing based on equation (2) allows enhancing accuracy of determination of the temperature and field dependences of magnetisation components from high-frequency EPR spectra. More details concerning described technique can be found in Ref. 17.

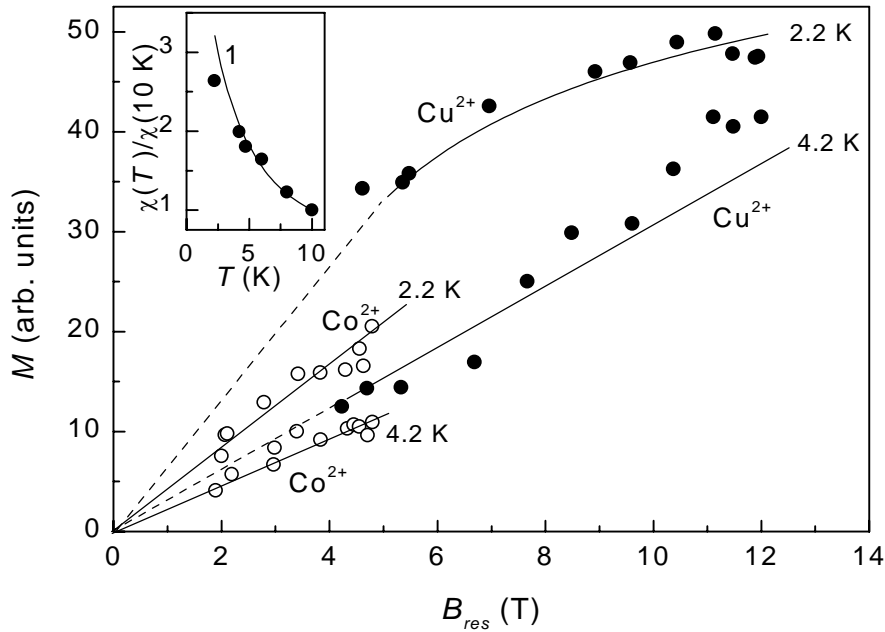

**Fig. 5** Field dependences of the various contributions into magnetisation of $CuGeO_3$:Co calculated from the EPR data at 4.2 K and 2.2 K. On the inset a comparison of the $\chi_{EPR}$ (points) with the static susceptibility data (curve 1) is presented.

Various magnetisation components restored from the spectroscopic data are presented in fig. 5. It is visible that for $Cu^{2+}$ chains the range $T \geq 4.2$K corresponds to linear magnetic response $M(B_{res},T) \sim B_{res}$ up to $B_{res} \sim 12$ T ($\omega/2\pi \sim 360$ GHz) whereas at $T=2.2$ K a pronounced tendency to saturation is already visible above $B_{res} \sim 6$ T ($\omega/2\pi \sim 150$ GHz). Simultaneously for all temperatures and frequencies studied the magnetisation of $Co^{2+}$ ions subsystem depends linearly on magnetic field (fig. 5).

The correctness of the aforementioned procedure of the separation of contributions into total magnetisation is illustrated by inset in fig. 5, where points denotes total spectroscopic susceptibility $\chi_{EPR} = M_1(T, B_{res1})/B_{res1} + M_2(T, B_{res2})/B_{res2}$ calculated for frequency $\omega/2\pi = 315$ GHz and curve 1 corresponds to magnetic susceptibility $\chi(T)$ measured for the same sample by vibrating sample magnetometer. One can see that data obtained by various methods agree well above helium temperature. For $T < 4.2$ K (see for example $T = 2$ K in fig. 5) the spectroscopic susceptibility $\chi_{EPR}(T)$ bends downwards with respect to $\chi(T)$ curve. This effect apparently originates from the saturation of the magnetisation for the $Cu^{2+}$ chains (fig. 5).

Therefore it is possible to conclude that the sum of contributions from EPR modes on $Cu^{2+}$ chains and $Co^{2+}$ ions adequately describes total magnetic susceptibility of $CuGeO_3$:Co sample. In this situation the magnetic subsystem of the $Co^{2+}$ ions define paramagnetic background $\chi_{Co^{2+}}(T)$ on which, according to [21], the spin-Peierls features with $T_{SP}(x=2\%)=12.5$ K should be observed. Staying in the frame of the universal doping scenario [13] it is reasonable to expect dropping of the magnetic susceptibility of the $Cu^{2+}$ chains $\chi_{Cu^{2+}}$ to zero in the range $T < T_{SP}(x=2\%) = 12.5$ K caused by opening of the spin gap. Exactly the same behavior of $\chi_{Cu^{2+}}$ has been obtained in [21] from the suggested in this work procedure of the subtraction of the paramagnetic "impurity" contribution from the total susceptibility $\chi(T) = \chi_{Cu^{2+}}(T) + \chi_{Co^{2+}}(T)$, which have allowed to "visualizing" spin-Peierls transition hardly noticeable on the strong paramagnetic background [21]. We wish to stress that according to interpretation of magnetic data suggested in [21] a condition $\chi_{Co^{2+}}(T) \gg \chi_{Cu^{2+}}(T)$ should be valid.

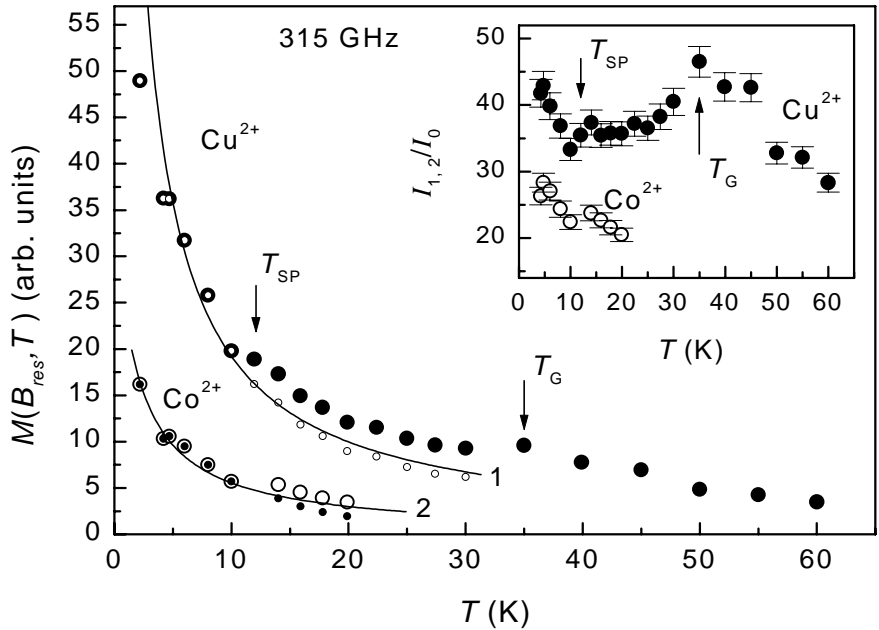

**Fig. 6** Temperature dependences of the various contributions into magnetisation of $CuGeO_3$:Co. Big points correspond to the $M(T)$ data and the small points denote result of the subtraction of the spin-Peierls contribution $M(T) - M_{SP}(T)$. On the inset the temperature dependences of the integrated intensities for the $Cu^{2+}$ chains and $Co^{2+}$ ions normalized on the integrated intensity of the DPPH resonance are presented. The figures near curves mark theoretical dependences: 1- Curie-Weiss law (equation (3)) or power dependence in the model of the quantum critical behavior (equation (4)); 2- Curie-Weiss law (equation (3)).

However the temperature dependences of the integrated intensities and magnetisations related with $Cu^{2+}$ chains and $Co^{2+}$ ions does not agree with assumptions proposed in [21] (see fig. 6). First of all, the integrated intensities for $Cu^{2+}$ and $Co^{2+}$ are comparable and the inequality $I_1>I_2$ is valid (inset in fig. 6). As a result the contribution to the total susceptibility of the sample from the $Cu^{2+}$ chains exceeds impurity contribution from $Co^{2+}$ ions (fig. 6).

Secondly, it is worth to note that on the temperature dependences of the ratio $I_1/I_0$ and magnetisation $M_1(B_{res},T)$ the feature that may be attributed to spin-Peierls transition is very weak. For comparison, we have marked by arrow in fig. 6 the temperature $T_{SP}=12$ K corresponding for $CuGeO_3$:Co sample with $x=2\%$ to a peak in specific heat registered in [21] and likely caused by spin-Peierls transition [21]. Our data show (fig. 6) that in vicinity of $T_{SP}=12$ K only weak, comparable with experimental error, decrease of $I_1$ and $M_1$ likely takes place and magnetic contribution of $Cu^{2+}$ chains for $T<T_{SP}$ does not turn to zero. Moreover, in the interval $2<T<30$ K a practically monotonous low temperature growth of $M_1(B_{res},T)$ is observed (рис. 6). As a result the temperature dependence of magnetisation of the $Cu^{2+}$ chains magnetic subsystem demonstrate noticeable deviations from the standard scenario.

It is interesting that both integrated intensity and magnetisation curves show more pronounced peculiarity at temperature $T=T_G=35$ K corresponding to onset of the low temperature enhancement of the $g$-factor (fig. 3). Around $T_G=35$ K the clear maximum of the ratio $I_1/I_0$ (inset in fig. 6) and related bend of the $M_1(T)$ curve for $Cu^{2+}$ chains (fig. 6) are clearly detected.

Therefore the set of experimental data obtained in the present work suggest that $CuGeO_3$ sample containing 2% of Co impurity acquires a new characteristic temperature $T_G=35$ K at which magnetic properties of $Cu^{2+}$ chains undergo considerable changes. In particular the range $T<T_G$ is characterized by strong damping of spin-Peierls transition accompanied by appearance of the quasi paramagnetic low temperature growth of magnetic moment (fig. 6).

We wish to point out that in contrast with [21] the use of high frequency EPR spectroscopy have allowed not only subtract impurity contribution, but also have opened an opportunity to separate correctly various contributions into magnetic properties of $CuGeO_3$:Co without any model *ad hoc* assumptions.

**4. Discussion.**

4.1 *Magnetic properties of* $Cu^{2+}$ *chains and possible quantum critical behavior.*

First off all, let us consider temperature dependence $M_1(T)$ for $Cu^{2+}$ chains (fig. 6). It is obvious that in case when all $Cu^{2+}$ chains in $CuGeO_3$:Co undergo spin-Peierls transition even with considerably reduced gap in magnetic excitation spectrum, the magnetic susceptibility for $T<T_{SP}$ could only decrease at low temperatures. Therefore a weak spin-Peierls anomaly at $T\sim12$ K, which is observed on a paramagnetic background, suggests an inhomogeneous distribution of cobalt impurity in the sample. As a result, for the majority of the chains the spin-Peierls transition becomes completely damped and for the minority of the chains the transition into dimerized state with reduced $T_{SP}=12$ K is conserved. The relative volume fraction of the chains with conserved spin-Peierls transition may be estimated from the magnitude of the kink on the temperature dependences $I_1(T)$ or $M_1(T)$, and according to fig. 6 should be about $\sim10\%$. This amount of the dimerized chains will be sufficient for the observation of the peculiarities at $T_{SP}$ in specific heat data [21].

As long as for the common spin-Peierls transition in $CuGeO_3$ the decrease of the EPR line width on $Cu^{2+}$ chains with lowering temperature and condition $g(T)\approx const$ are characteristic for both $T>T_{SP}$ and $T<T_{SP}$ [23], the low temperature growth of the $g$-factor (fig. 3) and broadening of the line width (fig. 4) reflect the characteristics of majority of the $Cu^{2+}$ chains with completely damped spin-Peierls state. In this situation the dimerization, which happens in a part of the chains, will mainly affect the temperature dependence of the integrated intensity that is observed experimentally (fig. 3,4,6).

For the quantitative analysis of the temperature dependence of magnetization for those $Cu^{2+}$ where spin-Peierls transition is completely damped we have corrected $M_1(T)$ data in the range $T<T_G=35$ K (fig. 6) by subtracting spin-Peierls part $M_{SP}(T)$ [28]. The resulting dependence $M_1(T)$-$M_{SP}(T)$ have been analyzed using Curie-Weiss law,

$$M(T) \sim \frac{1}{T-\Theta}, \qquad (3)$$

and also, following [16,17], using power law corresponding to the quantum critical behavior [18,19,29]

$$M(T) \sim \frac{1}{T^\alpha}, \qquad (4)$$

with the exponent $\alpha<1$. In our consideration only region of the linear magnetic response $T\geq 4.2$ K was analyzed for $Cu^{2+}$ chains.

It is found that for $Cu^{2+}$ chains both equations (3) and (4) allows describing well experimental data and curves representing best fits are practically identical (fig.6, curve 1). In the case of Curie-Weiss law (3) the estimated value of characteristic temperature $\Theta$ is $\Theta=-(0.8\pm 0.3)$ K, whereas fitting with power function have provided index $\alpha=0.93\pm 0.03$. Therefore the temperature dependence of magnetization of the major part of $Cu^{2+}$ chains with completely damped spin-Peierls transition is very close to Curie law.

The obtained result looks rather unexpected as long as in the absence of dimerization the magnetization of the quasi one-dimensional quantum spin S=1/2 antiferromagnetic chain should follow Bonner and Fisher law [30], for which, in contrast with Curie law, at low temperatures $T<J/k_B$ (where $J$ denotes absolute value of exchange integral in $Cu^{2+}$ chains) the magnetic susceptibility decreases with lowering temperature. Taking known from the literature [31] values $J/k_B$ for $CuGeO_3$ we get that susceptibility have to grow with temperature in the diapason $T<100$ K that contradicts to experiment (fig. 6). Consequently the observation of the $M(T)$ dependence close to Curie law may indicate that doping of $CuGeO_3$ with magnetic impurities induces *transition from quasi one-dimensional character of the spin chains to the three dimensional one*. In the latter case the doped $CuGeO_3$ should be treated as an anisotropic antiferromagnet described by Curie-Weiss law (3) with negative parameter $\Theta$. Indeed, as we have shown above, the best approximation of the experimental data $M_1(T)$-$M_{SP}(T)$ (fig. 6) is reached for $\Theta<0$.

In the frame of the proposed assumption it is interesting to estimate for $CuGeO_3$:Co a possible Neel temperature $T_N$. According to the theoretical calculations [27], when approaching $T_N$ in the paramagnetic area $T>T_N$ the width of the EPR line in an antifferomagnet diverges

$$w \sim \left(1-\frac{T_N}{T}\right)^{-3/4}. \qquad (5)$$

Equation (5) suggests that Neel temperature could be found from re-plotting of the experimental temperature dependence $w(T)$ in coordinates $w^{-4/3}=f(T^{-1})$ and than from extrapolation of the linear section to the value $w^{-4/3}=0$. It is visible from inset in fig. 4 that for $T\leq 12$K the curve $w^{-4/3}=f(T^{-1})$ really have a linear section corresponding to $T_N\approx 0.77$K. The estimated value is considerably smaller than Neel temperatures characteristic for the universal $T$-$x$ phase diagram of $CuGeO_3$. Indeed according to [2-12] for the concentration of impurity 2% the values $T_N\sim 2$-4 K are expected.

It is worth to consider another possible reason for "Curie-like" behavior of $M(T)$ in $CuGeO_3$:Co. Suppose that in spite of doping the *quasi one-dimensional character of $Cu^{2+}$ chains magnetic subsystem is conserved*. From the theoretical point of view [32] the susceptibility of one-dimensional chain with non-magnetic defects will follow Curie law if the condition $T<<4cJ/k_B$ is fulfilled (here $c$ stands for the relative concentration of defects in chain). Assuming for the estimate $c\sim x\sim 0.02$ and $J/k_B \sim 100$ K [31] we find that in our case $T<<8$ K, whereas in experiment the dependence close to Curie law is observed starting from $T\sim 30$ K (fig. 6). Therefore for $CuGeO_3$ the doping exactly by magnetic impurities is likely essential for appearance of the whole set of characteristic features of its physical properties including an anomalous temperature dependence of magnetisation of $Cu^{2+}$ chains.

In the quasi one-dimensional case the explanation of the data for $CuGeO_3$:Co (fig.1-6) could be given as well in the model of the disorder driven quantum critical behavior. [16-20,29]. Let us consider disorder in magnetic subsystem (caused, for example by doping), which is strong enough to decrease considerably temperature of the transition into magnetically ordered phase $T^*(x)$ or even to damp completely magnetic order ($T^*=0$). Than in certain temperature interval $T^*<T<T_G$ the magnetic subsystem will be in the Griffiths phase [18,19,29] that is characterized by the power dependence of magnetic susceptibility (see equation (4)). As it was mentioned above, the equation (4) also allows to describe well the low temperature growth of $M_1(T)-M_{SP}(T)$ for $Cu^{2+}$ chains (fig. 6).

Previously the quantum critical behavior in $CuGeO_3$ has been reported in [16,17,20]. It was found that doping of $CuGeO_3$ with magnetic impurity of iron on the concentration level $x\sim 0.01$ induces complete damping of the spin-Peierls state accompanied by onset of the power asymptotic of the magnetic susceptibility (equation (4)) with the critical exponent $\alpha\approx 0.35$. The transition temperature to the Griffiths phase have been estimated as $T_G\sim 20$-$70$ K [16,17,20]. Interesting that in contrast to $CuGeO_3$:Co in the case of $CuGeO_3$:Fe the single line caused by resonance on $Cu^{2+}$ chains have been observed in EPR spectra.

When comparing results of [16,17,20] with the data obtained in the present work, we wish to mark, first of all, that in the model of the disorder driven quantum critical behavior the parameter $\alpha$ is not universal and depends on the characteristics of the random fields in the sample [16-20,29]. As a consequence the magnitude of $\alpha$ for $CuGeO_3$ doped with different impurities may be essentially different and the observation of the exponent $\alpha=0.93$ in $CuGeO_3$:Co does not contradict to the quantum critical behavior model. Secondly, the characteristic temperature $T_G=35$K discovered in $CuGeO_3$:Co agrees with the estimate of the Griffiths temperature for $CuGeO_3$:Fe [16,17,20]. Therefore the model of the disorder driven quantum critical behavior allows explaining low temperature magnetisation data for $Cu^{2+}$ chains in $CuGeO_3$ (fig. 6) as well as peculiarities of the magnetic properties in the vicinity of $T_G$ attributing them to the transition to the Griffiths phase.

It is worth to note that quantum critical behavior and related power law for magnetisation (4) may appear in magnetic systems of various dimension having different types of magnetic interactions (see [16] and references cited therein). So, strictly speaking, the aforementioned arguments in favor of the possible observation of the quantum critical regime in cobalt doped $CuGeO_3$ does not lead to a definite conclusion about conservation of the quasi one-dimensional behavior of the $Cu^{2+}$ chains in this substance. Moreover the onset of the quantum critical region and transition into Griffiths phase does not exclude transition into magnetically ordered (for example antiferromagnetic) phase at lower temperatures, as long as for the observation of the quantum critical behavior the disorder induced decrease of the magnetic transition temperature is sufficient [29].

*4.2 The problem of the determination of the effective dimensionality of the $Cu^{2+}$ chains.*

It follows from the above consideration that consistent interpretation of the experimental data obtained in the present work requires determination of the effective dimensionality of the $Cu^{2+}$ chains magnetic subsystem. Apparently the information presented in fig. 5-6 is insufficient for the solution of this problem. Therefore the possibilities originating from the existing theoretical models for the temperature dependences of the line width and *g*-factor will be considered below (fig. 3,4).

Recently Oshikawa and Affleck have suggested a new theory of EPR for one-dimensional homogeneous antiferromagnetic chains with spin S=1/2 [24]. According to their approach the presence of the staggered field in such spin system leads to the simultaneous low temperature growth of the line width and *g*-factor. Moreover, as it was shown in [33] the line width $w(T)$ and temperature dependent correction to the *g*-factor $\Delta g(T)$ are linked by the universal relation

$$\frac{w(T)}{\Delta g(T)} = 1.99 \frac{k_B T}{\mu_B}, \qquad (6)$$

which does not depend on the magnitude of the staggered field and exchange integral. This consequence of the Oshikawa and Affleck theory appears to be unique in the EPR theory and is characteristic just for the one-dimensional case [24,33]. As a result the comparison of the experimental data with the theoretical dependence (6) may serve as an additional argument for the determination of the magnetic subsystem dimension [33].

The specific character of the predicted in [24,33] behavior clearly reveal itself in the estimate of the ratio $w(T)/\Delta g(T)$ in the model of the three-dimensional antiferromagnet. In this case the $w(T)$ in paramagnetic region is given by equation (5) (see also [27]) and the changes in the *g*-factor may be attributed to the difference between local field (which affects spins directly) and the external one, i.e. we expect $\Delta g(T) \sim \chi(T)$. The latter assumption is widely used for the interpretation of the EPR data in spin glasses [34]. Therefore for the three-dimensional antiferromagnet it is possible to write

$$\frac{w(T)}{\Delta g(T)} = A \frac{T^{3/4}(T + |\Theta|)}{(T - T_N)^{3/4}}, \quad T > T_N, \qquad (7)$$

where coefficient *A* depends on the coupling constant between susceptibility and correction to the *g*-factor. Equation (7) suggests that in three-dimensional case the ratio diverges: $w(T)/\Delta g(T) \to \infty$, when $T \to T_N$, whereas in the Oshikawa and Affleck theory for the one-dimensional case $w(T)/\Delta g(T) \to 0$ for $T \to 0$ (equation (6)).

As follows from the aforementioned experimental results, in $CuGeO_3$:Co the anomalous low temperature growth of the resonance line width and *g*-factor is observed just on the $Cu^{2+}$ chains where dimerization, which is not considered in the Oshikawa and Affleck theory, is completely damped (see fig. 4 and fig. 3 as well as discussion in section 4.1). Simultaneously on the curves $M(T)$ for $Cu^{2+}$ chains (fig. 6) in the diapason $T > 2$ K any features characteristic to antiferromagnetic ordering are missing (see also the estimate of the Neel temperature from the temperature dependence of the line width (inset in fig. 4)). Therefore equations (6) and (7) may be applied for the analysis of the experimental data in fig. 3,4.

The ratio $w(T)/\Delta g(T)$ have been calculated assuming $\Delta g(T)=g(T)-g(30 K)$ (see inset in fig. 3) and the values $w(T)$ for $Cu^{2+}$ chains have been taken from fig. 4. The obtained result for $T < 20$ K is presented in fig. 7 by black circles. As long as the similar low temperature anomalies of the temperature dependences of the line width and *g*-factor have been observed in $CuGeO_3$:Fe [16], on the same figure we have plotted the values $w(T)/\Delta g(T)$ for this compound (white circles in fig. 7), which have been calculated in Ref. 33. It is worth to note that in spite of considerable difference in parameters of the EPR line for $Cu^{2+}$ chains in $CuGeO_3$:Co and $CuGeO_3$:Fe [33], the magnitude $w(T)/\Delta g(T)$ for these compounds practically coincides within experimental error. Additionally the Oshikawa and Affleck theory (equation (6)) predicts correctly the absolute value and character of the temperature dependence of $w(T)/\Delta g(T)$ in $CuGeO_3$, doped with magnetic impurities (fig. 7, curve 1). However it follows from fig. 7 that calculation [24] overestimates somehow the value of the numerical coefficient in equation (6). Best fit of the experimental data with the help of linear dependence gives $w(T)/\Delta g(T)=(1.63\pm0.07)k_BT/\mu_B$ (fig. 7, curve 2) and empirical magnitude of the coefficient in equation (6) becomes 20% less than the theoretical one. Nevertheless, when taking into account, first of all, the accuracy and character of assumption used in the Oshikawa and Affleck theory [24,33] and, secondly the fact that [24,33] in $CuGeO_3$ the interchain exchange has considerable amplitude [31], the agreement between theory and experiment can be called more than satisfactory.

As long as criterion $w(T)/\Delta g(T)$ in the Oshikawa and Affleck theory does not contain any free parameters (equation (6)) the above analysis provides a serious argument in favor of the conservation of the one-dimensional character of $Cu^{2+}$ spin chains in condition of damping of the spin-Peierls transition. At the same time, if we use the values of $\Theta$ and $T_N$ found in section 4.1 and consider the coefficient $A$ in equation (7) as fitting parameter the reasonable agreement of the ratio $w(T)/\Delta g(T)$ calculated in the model of three-dimensional antiferromagnet with the experimental data for $CuGeO_3$:Co at $T>2$ K can be reached also (fig. 7, curve 3). It is clearly visible from fig. 7 that the difference between curves 1,2 and curve 3 becomes most pronounced in the temperature diapason $T<1$ K. Therefore the investigation of the EPR at very low temperatures in $CuGeO_3$ doped with magnetic impurities could be rewarding and will be the subject of our future studies.

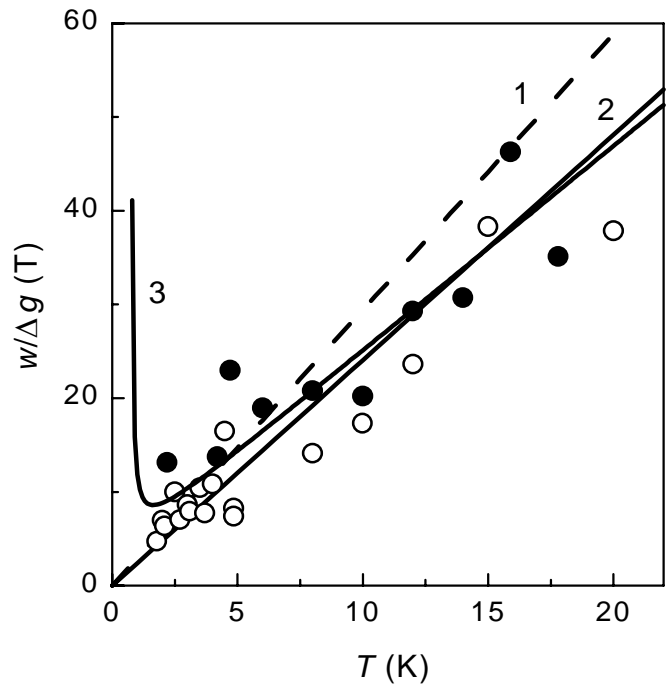

**Fig. 7** The function $w(T)/\Delta g(T)$. Black points denote experimental data for $CuGeO_3$:Co and white points correspond to experimental data from [33] for $CuGeO_3$:Fe. 1- the universal dependence (6) in the Oshikawa and Affleck theory (one-dimensional case); 2- best fit of the experimental data with the linear dependence; 3- the model of the three-dimensional antiferromagnet (equation (7)).

4.3 *Possible EPR on* $Co^{2+}$ *ions in* $CuGeO_3$ *matrix.*

In above consideration we have suggested that observed in $CuGeO_3$:Co resonance 2 (fig. 1,a) is caused by EPR absorption on $Co^{2+}$ ions rather than by antiferromagnetic resonance (section 3.1). Apparently this line in the EPR spectrum cannot be attributed to harmonic of the main resonance on $Cu^{2+}$ chains. This assumption is not supported by frequency dependence of the resonance 2 (fig. 1, b) as well as by the fact that in contrast to the resonance 1 the line width increases with temperature (fig. 4). More exotic explanation based, for instance, on the predicted in the Oshikawa and Affleck theory soliton type and breather type excitations [24] are also look unlikely. [24]. Indeed in this situation, first of all, the main resonance on $Cu^{2+}$ should be damped and, secondly, the dispersion curve $\omega_2(B)$ must be essentially non-linear [24]. Both predictions do not meet experimental data in the studied temperature and frequency domain (fig. 1,2,6).

However the explanation of the resonance 2 by EPR mode on $Co^{2+}$ ions in $CuGeO_3$ matrix is rather nontrivial as long as up to now individual modes caused by impurities doping $Cu^{2+}$ chains in $CuGeO_3$ and modifying their properties have not been known. Probably the appearance of such a features in magneto-optical spectra is a consequence of inhomogeneous distribution of impurities in the sample, likely characteristic for $CuGeO_3$:Co system. As an indication on this possibility may serve results of analysis of the $M(T)$ data for $Cu^{2+}$ chains (fig. 6), which demonstrate existence of the two types of $Cu^{2+}$ chains: the first kind (volume fraction ~10 %) with spin-Peierls transition and the second kind (volume fraction ~90%) with complete damping of the spin-Peierls transition (see. section 4.1). This difference can be easily explained by inhomogeneous doping of the spin chains in $CuGeO_3$:Co.

As long as the mentioned anomalies are observed in the concentration range below the solubility limit of Co impurity in $CuGeO_3$ matrix [21], it is reasonable to connect them with the strong tendency to clusterization of Co atoms. It is also possible to assume that in the spatial regions where Co clusters are formed, the contents of the Co impurity in $Cu^{2+}$ chains will decrease and for these chains the spin-Peierls transition may be conserved. Simultaneously we may expect that ions inside cluster will provide an independent contribution into magnetic properties of the sample, which will manifest itself, in particular, by additional EPR line (fig. 1). As a result, the renormalization of the $g$-factor for $Co^{2+}$ (fig. 3) may be caused by interaction effects inside cluster (see point 3.1) or attributed to the influence of the $Cu^{2+}$ chains surrounding cluster on its properties. The feature on the curves $I_2(T)$ and $M_2(T)$ observed in the vicinity of $T_{SP}$ (fig. 6) favors the latter explanation. Therefore the nature of the "EPR on $Co^{2+}$ ions in $CuGeO_3$ matrix" may be rather complicated and the full understanding of the characteristic of this line will probably require additional theoretical investigations.

As in the case of the resonance on $Cu^{2+}$ chains for the considered feature a quantitative analysis of the magnetisation temperature dependence have been carried out (fig. 6). We have used similar procedure for subtraction of the kink at $T_{SP}$=12 K after which the $M(T)$ data for the resonance 2 was analyzed by Curie-Weiss law (equation (3)) and power dependence (equation (4)). It is found that the using of equation (4) for fitting leads to enhancement of the square error of approximation by 1.5 times with respect to using of the equation (4). The best fit was obtained for Curie-Weiss law with characteristic temperature $\Theta$= -(1.8±0.5) K (curve 2 in fig. 6). Interesting that in the considered case the absolute value of $\Theta$ is essentially bigger than for the $Cu^{2+}$ chains where $\Theta$~ -0.8 K have been estimated. Therefore the three-dimensional antiferromagnetic correlations in $Co^{2+}$ ions subsystem in $CuGeO_3$:Co are likely more pronounced than in subsystem of the $Cu^{2+}$ chains. The obtained result underlines once again the different physical nature of the contributions controlling magnetic properties of $CuGeO_3$:Co and related EPR features.

## 5. Conclusion.

Our study of the resonant magnetoabsorption spectra in germanium cuprate shows that doping of this compound by 2% of cobalt induces a set of unusual and unknown before for doped $CuGeO_3$ peculiarities in its physical properties.

First of all, it is necessary to note the appearance of a new line of the EPR absorption (resonance 2 in fig. 1,2), which is essentially different from the resonance on antiferromagnetic chains of cooper ions. This feature is characterized by linear dependence of the resonant frequency on magnetic field corresponding to $g$-factor with the magnitude $g\approx4.7$ that is more than two times bigger than $g$-factor for $Cu^{2+}$ chains where $g\approx2.15$. Analysis of the line shape for both resonances shows that there is a qualitative difference in the temperature dependences of the line widths for the resonances forming EPR spectrum in CuGeO$_3$:Co. Whereas for the resonance 2 the line width decreases with lowering temperature in agreement with the standard theories of spin relaxation, the line width for the EPR on Cu$^{2+}$ chains increases three times when temperature is lowered from 60 K to 4.2 K. Therefore the doping with cobalt not only leads to appearance of the new line in spectra but also modifies considerably the properties of the Cu$^{2+}$ chains inducing an anomalous low temperature growth of the line width (note that for CuGeO$_3$ the decrease of the line width with lowering temperature is observed generally for this resonance [2,7,23]).

Considering the obtained results from the point of view of the universal $T$-$x$ phase diagram [1-15] we wish to emphasize that temperature dependence of magnetisation for the Cu$^{2+}$ chains $M_1(T)$ calculated from the integrated intensity for the corresponding resonance (resonance 1 in fig. 1,2) exhibits two characteristic features, namely known before spin-Peierls transition at $T_{SP}$=12 K [21] and a kink of $M_1(T)$ at $T_G$ =35 K. In the region $T<T_G$ additionally to the peculiarity in magnetization the onset of the low temperature growth of the $g$-factor for the resonance on Cu$^{2+}$ is observed. This provides an independent confirmation of the appearance in CuGeO$_3$:Co of a new characteristic temperature $T_G$, which is essentially bigger than all transition temperatures known for the universal $T$-$x$ phase diagram.

For the explanation of the obtained data it is necessary to assume that likely the spin-Peierls transition takes place for approximately 10% of Cu$^{2+}$ chains whereas in the rest of the sample volume (90% of chains) the spin-Peierls state is completely destroyed by doping with cobalt. This unusual behavior may be caused by spatially inhomogeneous distribution of impurity in the sample leading to formation of the clusters from cobalt ions in germanium cuprate matrix. If follows from the consideration of the known up to now theoretical possibilities that forming of the clusters consisting of Co$^{2+}$ magnetic ions is the most probable reason for appearance of the new absorption line in the EPR spectra (fig. 1,2).

The problem of correct subtraction on the impurity paramagnetic background, which may mask expected in theory for one-dimensional case the $M(T)$ temperature dependence of Bonner-Fisher type, is of great practical importance for analysis of the magnetic properties of low dimensional magnets. In the present work this problem have been solved experimentally without involving any models or assumption concerning paramagnetic background. We have shown that in conditions of damping of the spin-Peierls states for majority of Cu$^{2+}$ chains the temperature dependence of magnetization is very close to Curie law. Therefore the paramagnetic character of magnetic susceptibility in CuGeO$_3$:Co does not related with paramagnetic impurities as it was suggested earlier [21], but mainly reflects intrinsic properties of doped Cu$^{2+}$ chains.

The observed deviations from the universal scenario of doping, which includes absence of antiferromagnetic transitions in magnetic subsystems of Cu$^{2+}$ chains and Co$^{2+}$ ions for the range $T>2$ K, demand searching for alternative theoretical approaches. For this purpose we have performed a comparative analysis of experimental data in the frame of the models of (i) quantum critical behavior accounting the Oshikawa and Affleck theory for one-dimensional systems and (ii) three-dimensional antiferromagnet with reduced by disorder Neel temperature. It is established that both approaches provide quantitative explanation of experimental data for the temperature range $T>2$ K. However in our opinion the model of quantum critical behavior looks more reliable, as long as it allows to explain naturally the characteristic temperature $T_G$ =35 K by transition into Griffiths phase. The final choice of the most adequate approach requires EPR studies of CuGeO$_3$:Co at very low temperatures where the difference between two models will become apparent.

**Acknowledgements.**

The work have been supported by grants RFBR 04-02-16574, INTAS 03-51-3036 and grant PD02-1.2-335 of the Russian Ministry of Education. Some aspects of the present study have been supported by programmes "Low dimensional quantum structures", "Strongly correlated electrons" of Russian Academy of Sciences and "Integration" of Russian Ministry of Education. SVD acknowledge support from Russian Science Support Foundation and Venture Business Laboratory (Kobe University).